\documentclass[10pt]{article}
\usepackage{epsfig}
\title{3-dimensional scalar-vector dual of topological $\sigma$-model\thanks{%
Talk delivered at the {\it EuroConference on Partial Differential Equations and their Applications to Geometry \& Physics} in Castelvecchio Pascoli.}}

\author{Bogus\l aw Broda\\
\it Department of Theoretical Physics\\
\it University of \L\'od\'z\\
\it Pomorska 149/153\\
\it PL--90-236 \L\'od\'z\\
\it Poland}

\begin{document}
\bibliographystyle{unsrt}
\nocite{*} 

\maketitle

\begin{abstract}
A 3-dimensional model dual to the Rozansky-Witten topological sigma-model with a hyper-K\"ahler target space is considered. It is demonstrated that a Feynman diagram calculation of the classical part of its partition function yields the Milnor linking number.
\end{abstract}

\section{Introduction}
The first milestone in the development of what is called now
topological quantum field theory (i.e.\ application of quantum field theory to low-dimensional topology) are two papers by Witten. The first one, describing Donaldson's invariants of 4-dimensional manifolds using BRST-like cohomology \cite{Witten:1988ze}. The second one, deriving the Jones invariant of knots and links and its generalizations as well as the Reshetikhin-Turaev-Witten invariants of 3-dimensional manifolds, using 3-dimensio\-nal
Chern-Simons gauge field theory \cite{Witten:1989hf}. The second milestone are the papers by Seiberg and Witten on dual, low-energy version of Donaldson-Witten theory \cite{Seiberg:1994rs} and \cite{Witten:1994cg}. Upon dimensional reduction the 4-dimensional
theories can serve as a source of new 3-dimensional topological field theories which, in turn, can provide in low energy further
topological field theory models. These latter 3-dimensional models belong to
the class of topological $\sigma$-models proposed recently by Rozansky and Witten \cite{Rozansky:1997bq}. In this way, we have the two alternative (and perhaps complementary) topological field theories probing topology of 3-dimensional manifolds: Chern-Simons theory and Rozansky-Witten theory. In fact, Rozansky-Witten (RW) theory is
similar to perturbative Chern-Simons theory since Feynman diagrams are
analogous in the both theories as well as resulting topological invariants. Further study of the RW invariants and their relations to other topological invariants are presented
in \cite{Habegger:1999yp}.

Inspired by the paper of Rozansky and Witten \cite{Rozansky:1997bq} we aim to continue
this line of research. In the original RW model
we have a multiplet of scalar fields assuming values on a hyper-K\"ahler
manifold $X$ as a target space. But for a 3-dimensional manifold
$M$ and a hyper-K\"ahler manifold $X$ with symmetries we can
alternatively switch to a dual description where the scalar fields
are replaced by Abelian vector gauge fields \cite{Hitchin:1987ea}. Actually, we are interested in the simplest possibility for this approach, namely
4-component multiplet of scalar fields ($\dim_RX=4$) with one scalar field
dualized to a single vector field.
In the paper \cite{Broda:1999zw}, topological invariants for the target space of the particular form $X=S^1\times X'$ have been derived. In
the present paper, no this particular form of the target space is assumed. Instead we will show that for a hyper-K\"ahler target
space $X$ with a Killing vector, for which one scalar field is
dualized to an Abelian vector gauge field, the ``classical'' part
(the part of the zero order in the constant $\hbar$) of the partition function is easily
calculable, and corresponds to the whole partition function of the RW model
for the first Betti number $b_1(M)=2$. The partition function is expressible by the Reidemeister-Ray-Singer torsion and the Milnor linking number. We should stress that no reference to topological (BRST-like) charges is used.

\section{Topological model}
The starting point of our analysis is the RW model
\cite{Rozansky:1997bq}, which is a topological quantum
$\sigma$-model on a 3-dimensional manifold $M$, parameterized by 4-dimensional (possibly, real $4n$-dimensional) hyper-K\"ahler manifolds
$X$. It is related to ${\cal N}=4$ SUSY $\sigma$-model via the,
so-called, twist. The action for this model is of the following
form
\begin{eqnarray}
    S=S_{\rm B}+S_{\rm F}
    &=&\int_M\left[\left(\frac{1}{2}\sqrt{h}\,g_{ij}
    \partial_\mu\phi^i\partial^\mu\phi^j\right)
    +\left(\sqrt{h}\,\epsilon_{IJ}
    \chi_\mu^I\nabla^\mu\eta^J\right.\right.
    \label{SB+SF}\\
    &&\!\!\!\left.\left.+\frac{1}{2}\epsilon^{\mu\nu\lambda}
    \epsilon_{IJ}\chi_\mu^I\nabla_\nu\chi_\lambda^J
    +\frac{1}{6}\epsilon^{\mu\nu\lambda}\Omega_{IJKL}
    \chi_\mu^I\chi_\nu^J\chi_\lambda^K\eta^L
    \right)\right]d^3x,
    \nonumber
\end{eqnarray}
where $h_{\mu\nu}(x)$ and $g_{ij}(\phi)$ are metric tensors on $M$
and $X$ respectively, the fermion scalars $\eta^I$ and the fermion
1-forms $\chi_\mu^I$ assume values in the rank 2 complex vector
bundle $W$ originating from the decomposition of the
complexification of the tangent bundle $TX$ of $X$.
$\Omega_{IJKL}$ corresponds to the Riemann tensor on $X$, and the
covariant derivative is defined using the Levi-Civita connection
on $M$ and $W$ (see \cite{Rozansky:1997bq} for details). Thus,
$\mu,\nu,\lambda=1,2,3$, $i,j=1,2,3,4$, and $I,J,K,L=1,2$.
``Topologicallity'' of this theory is confirmed a priori by the
existence of a nilpotent BRST charge (in fact, 2 charges), and a
posteriori by the derivation of the Casson-Walker-Lescop
invariant.

Now, let as assume that the target space $X$ has a continuous
internal symmetry, i.e. the hyper-K\"ahler manifold $X$ has a
Killing vector. Then we can perform a duality transformation of
the action (\ref{SB+SF}) replacing one scalar field (say, $\phi^{(4)}$)
by an Abelian gauge 1-form $A_\mu$ \cite{Hitchin:1987ea}. The
result of such a transformation in the bosonic sector $S_{\rm B}$
provides a modification, 
\begin{eqnarray}
    S_{\rm B}\longmapsto S_{\rm B}'
    &=&\int_M\left(\frac{1}{2}\sqrt{h}\,\tilde{g}_{ab}(\phi^c)
    \partial_\mu\phi^a\partial^\mu\phi^b\right.
    \label{SB'}\\
    &&\left.+\frac{1}{4}\sqrt{h}\,g^{-2}(\phi^c)
    F_{\mu\nu}F^{\mu\nu}
    +i\epsilon^{\mu\nu\lambda}g_a(\phi^c)F_{\mu\nu}
    \partial_\lambda\phi^a\right)d^3x,
    \nonumber
\end{eqnarray}
where $F_{\mu\nu}=\partial_\mu A_\nu-\partial_\nu A_\mu$, and
$\tilde{g}_{ab}$, $g_a$ and $g$ are functions of $\phi^c$ only with $a,b,c=1,2,3$, i.e.
$$
\tilde{g}_{ab}=g_{ab}-(g_{00})^{-1}g_{0a}g_{0b},
\qquad
g_a=g_{0a},
\qquad
g^2=g_{00}.
$$
Consequently, in quantum theory one should also add appropriate gauge-fixing and Faddeev-Popov terms.

\section{The one-loop calculus}
We conclude that the partition function of our model is expressed by the following path integral
\begin{equation}
    Z(M)=\int D\phi^a DA_\mu D\eta^I D\chi_\mu^I
    D\bar{\theta} D\theta\,
    \exp{\left(-\frac{1}{\hbar}S\right)},
    \label{Z(M)}
\end{equation}
where the action
\begin{equation}
    S=S_{\rm B}'+S_{\rm F}+S_{\rm gf}+S_{\rm FP}
    \label{S}
\end{equation}
includes also the gauge-fixing part $S_{\rm gf}$ and
Faddeev-Popov term $S_{\rm FP}$.
Now, we will identify minima of the action and expand the boson
fields around them. The minima of the action consist of pairs
$(\varphi_{\rm o}^a ,A_{{\rm o}\mu})$, where $\varphi_{\rm o}^a$ is a constant map, and $A_{{\rm o}\mu}$ is a flat connection on $M$.
Having an expansion point, we are able to explicitly express the
gauge-fixing and Faddeev-Popov terms. Namely,
\begin{equation}
    S_{\rm gf}+S_{\rm FP}=\int_M g^{-2}(\varphi_{\rm o})
    \left   (\frac{1}{2}\nabla^\mu A_\mu\nabla^\nu A_\nu
    +\partial_\mu\bar{\theta}\partial^\mu\theta\right)
    \sqrt{h}\,d^3x,
    \label{S_gauge}
\end{equation}
with the Faddeev-Popov ghost fields $\bar{\theta}$, $\theta$, and $\varphi_{\rm o}$ a constant $\sigma$-field.

There is a host of zero modes present in our model (see, table \ref{0-modes}),
which are postponed to higher-loop calculus.
Bosonic scalar zero modes correspond to constant maps
$\varphi_{\rm o}^a$, and their number, the dimension of the moduli space
of constant maps of $M$ to $\tilde{X}$, equals the dimension of
the (reduced) target manifold $\dim\tilde{X}=3$ (more generally,
it equals $b_0(M)\dim\tilde{X}$). Bosonic vector zero modes are
tangent to the moduli space of flat connections $A_{{\rm o}\mu}$, and
their number follows from the de~Rham cohomology, and is equal to
the first Betti number $b_1(M)$. The moduli space of flat
connections is a torus of dimension $b_1(M)$. Likewise the numbers
of fermionic zero modes depend on Betti numbers $b_0$ and $b_1$
for scalar $\eta^I$ and vector $\chi_\mu^I$ fields, respectively. The single zero mode
of the ghost fields should be removed from the beginning as it
corresponds to a trivial gauge transformation of $A_\mu$.

%
\begin{table} %
\centerline{\hbox{\begin{tabular}{ccc}
\hline
{\bf TYPE}          & {\bf FIELD}                               & {\bf NO.} \\
    \hline
    \hline
    BOSONIC ($+$)     \\
    \hline
scalar          & $\phi^a$                           & 3             \\
vector              & $A_\mu$                               & $b_1$         \\
    \hline
    FERMIONIC ($-$) \\
    \hline
scalar (ghost)  & $\bar{\theta}$, $\theta$  & 1                 \\
    \hline
scalar                  &   $\eta^I$                                    &   2                   \\
vector                  & $\chi_\mu^I$                          & $2b_1$        \\
    \hline
\end{tabular}}}
\caption[]{Zero modes present in the theory and their respective
numbers.}\label{0-modes}
\end{table}

Following \cite{Rozansky:1997bq}, we split the bosonic scalar
field $\phi^a(x)$ into an orthogonal sum
\begin{equation}
    \phi^a(x)=\varphi_{\rm o}^a+\varphi^a(x),
\end{equation}
where $\varphi_{\rm o}^a$ is an expansion point and $\varphi^a(x)$ represents a non-constant (fluctuating) part
of the $\phi$ field. The partition function can now be expressed
as an integral over $\tilde{X}$
\begin{equation}
    Z(M)=(2\pi\hbar)^{-\frac{3+b_1-1}{2}}
    \int_{\tilde{X}} V\big(T^{b_1}\big)
    Z(M;\varphi_{\rm o}^a)\sqrt{g}\,d^3\varphi_{\rm o}.
    \label{Z_X(M)}
\end{equation}
where the power of the (Planck) constant $\hbar$ follows from counting of the zero
modes given in table \ref{0-modes}
(the numbers in the numerator of the exponent equal the number of ``boson''$-$``ghost'' zero modes), and $V\big(T^{b_1}\big)$ is
the volume of the torus of classical minima which is independent of $\hbar$ \cite{Witten:1995gf}. $Z(M;\varphi_{\rm o}^a)$
is a contribution coming from the one-loop zero
mode-free part $Z_0(M)$, and a higher-order part $Z_I(M)$ which should
saturate fermionic zero modes.

The one-loop part $Z_0$ is given by
functional determinants of non-zero modes of differential
operators entering the free part of the action in the expansion
around $\varphi_{\rm o}$. The differential operators in question are:
 \begin{itemize}
\item
Laplacians acting on 0-forms $\Delta_0$ for the scalar field
$\varphi$ and for the Faddeev-Popov ghost fields $\bar{\theta}$, $\theta$;
\item
Laplacian acting on 1-forms $\Delta_1$ for the vector gauge field $A_\mu$;
\item
The differential operator \cite{Rozansky:1997bq}
\begin{equation}
    L_-(\eta,\chi_\mu)
    =\left(-\nabla^\mu\chi_\mu,\nabla_\mu\eta
    +h_{\mu\nu}\frac{1}{\sqrt{h}}\epsilon^{\nu\rho\lambda}
    \partial_\rho\chi_\lambda\right),
\end{equation}
for non-ghost fermionic fields $\eta$ and $\chi_\mu$.
    \end{itemize}
Collecting all the determinants we obtain
\begin{equation}
    Z_0(M)=\frac{\det'L_-}%
    {\left[\det'(-\Delta_0)\det'(-\Delta_1)\right]^{1/2}},
    \label{Z_0}
\end{equation}
where the primes mean discarding zero modes. It appears that the absolute value of the ratio of the determinants in (\ref{Z_0}) is
related to the Reidemeister-Ray-Singer analytic torsion
$\tau_R(M)$ of the trivial connection on $M$, i.e.
\begin{equation}
    \left|\frac{\det'L_-}{\left[\det'(-\Delta_0)
    \det'(-\Delta_1)\right]^{1/2}}\right|
    =\tau_R^{-2}(M).
\end{equation}

\section{Beyond one loop}
First of all, we should determine the form of propagators
entering our model. According to (\ref{S}), upon expansion
around the constant field $\varphi_{\rm o}$ we obtain the following
quadratic part of the action
\begin{eqnarray}
    S_0&=&\int_M \sqrt{h} \left(\frac{1}{2}\tilde{g}_{ab}
    \partial_\mu\varphi^a\partial^\mu\varphi^b
    +\frac{1}{2}g^{-2}
    \nabla^\mu A_\mu\nabla^\nu A_\nu\right.
    \label{S0}\\
    &&\left.+\epsilon_{IJ}\chi_\mu^I\nabla^\mu\eta^J
    +\frac{1}{2}h^{-1/2}\epsilon^{\mu\nu\lambda}\epsilon_{IJ}
    \chi_\mu^I\nabla_\nu\chi_\lambda^J\right)d^3x.
    \nonumber
\end{eqnarray}
The ghost part is absent in $S_0$ because it was
integrated out. Similarly, the topological part
$g_a\epsilon^{\mu\nu\lambda} F_{\mu\nu}\partial_\lambda\varphi^a$
does not enter (\ref{S0}). The fermionic part of the action (\ref{S0})
can be expressed using the operator $L_-$ \cite{Rozansky:1997bq}
as
\begin{equation}
    \frac{1}{2}\epsilon_{IJ}\left<\eta^I,\chi_\mu^I\left|L_-
    \right|\eta^J,\chi_\nu^J\right>.
    \label{L-}
\end{equation}
Therefore, the non-zero propagators assume the
following form:
\begin{eqnarray}
    \left<\varphi^a(x)\varphi^b(y)\right>
    &=&-\hbar\tilde{g}^{ab}G_0'(x,y),
    \nonumber\\
    \left<A_\mu(x)A_\nu(y)\right>
    &=&-\hbar h_{\mu\nu}G_1'(x,y),
    \nonumber\\
    \left<\chi_\mu^I(x)\chi_\nu^J(y)\right>
    &=&\hbar\epsilon^{IJ}G_{\mu\nu}'(x,y),
    \nonumber\\
    \left<\eta^I(x)\chi_\mu^J(y)\right>
    &=&\hbar\epsilon^{IJ}\partial_\mu^x G_0'(x,y),
    \nonumber
\end{eqnarray}
where $G$'s denote Green's functions and the prime means the absence of zero modes. 

\begin{figure}[ht]
\centering
\epsfig{file=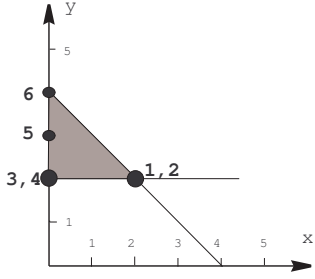, width=5.5cm}
\caption{Solutions of the system (\ref{4,y}).} \label{Fig:solutions}
\end{figure}

\begin{figure}[ht]
\centering
\epsfig{file=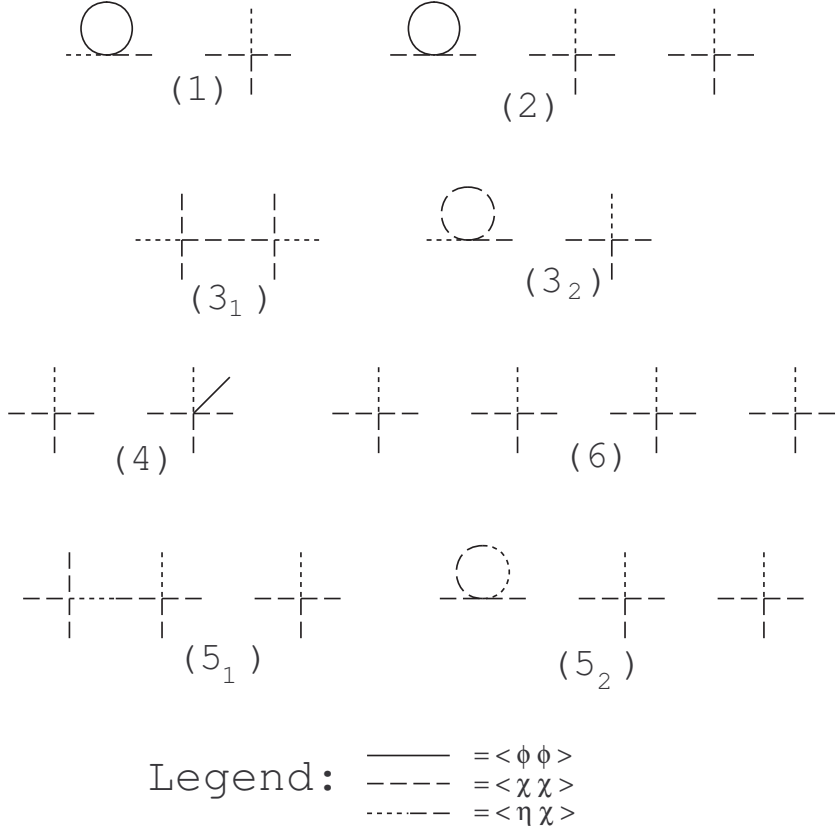, width=11.5cm}
\caption{Feynman diagrams for the ``classical'' part of $Z(M)$.} \label{fig:Diagrams}
\end{figure}

Now, we should determine the class of Feynman diagrams which could
play a role in our further analysis. The condition selecting only classical part, i.e.\ the terms canceling the power of the Planck constant $\hbar$ in front of (\ref{Z_X(M)}) is very severe. Following the arguments of
Rozansky and Witten \cite{Rozansky:1997bq} we preselect the
candidate Feynman diagrams as those of the order
$\hbar^{\frac{3+b_1-1}{2}}=\hbar^{\frac{2+b_1}{2}}$, to cancel the normalization factor in (\ref{Z_X(M)}). Therefore, if $V$ is the number of vertices,
and $L$ is the total number of legs emanating from all of the various
vertices, then for our diagrams we should have
\begin{equation}
    \frac{L}{2}-V=\frac{2+b_1}{2}.
    \label{L/2}
\end{equation}
Since interaction vertices are of at least of the fourth order, we require
\begin{equation}
    L\geq 4(V_0+V_1+V_2+V_3),
    \label{L}
\end{equation}
where $V_n$ ($n=0,1,2,3$) denote vertices with $n$ $\chi$ fields.
Finally, to absorb $2b_1$ $\chi$ zero modes, we need
\begin{equation}
    V_1+2V_2+3V_3\geq 2b_1,
    \label{2b1}
\end{equation}
whereas to absorb 2 $\eta$ zero modes
\begin{equation}
    V_1+V_3\geq 2.
    \label{2}
\end{equation}
From (\ref{L/2}) and (\ref{L}) it follows that
\begin{equation}
    2+b_1\geq 2(V_0+V_1+V_2+V_3).
    \label{2+b1}
\end{equation}
Inserting (\ref{2b1}) into (\ref{2+b1}) yields
\begin{equation}
    4\geq 4V_0+3V_1+2V_2+V_3.
    \label{4}
\end{equation}
Introducing the notation
\begin{eqnarray}
    x&=&4V_0+2V_1+2V_2
    \nonumber\\
    y&=&V_1+V_3,
    \label{x,y}
\end{eqnarray}
we can rewrite (\ref{4}) and (\ref{2}) as
\begin{eqnarray}
    4&\geq&x+y
    \nonumber\\
    y&\geq&2.
    \label{4,y}
\end{eqnarray}
The solutions of (\ref{4,y}) denoted with dots are given in
figure \ref{Fig:solutions}, where according to (\ref{x,y})
$x$ assumes only even values.
%
%
The four solutions of the system of inequalities (\ref{4,y}) give
rise to the following set of six solutions to the primary system (\ref{L/2}), (\ref{L}), (\ref{2b1}), (\ref{2}) presented in table \ref{solutions}.
%
\begin{table} %
\centerline{\hbox{
\begin{tabular}{ccccccccc}\hline
{~~\bf No.~~}&{$V_0$}&{$V_1$}&{$V_2$}&
{$V_3$}&{$b_1(M)$}&{$L$}&{$x$}&{$y$}\\\hline
\bf(1)&0&1&0&1&2& 8&2&2\\
\bf(2)&0&0&1&2&4&12&2&2\\
\bf(3)&0&0&0&2&2& 8&0&2\\
\bf(4)&0&0&0&2&3& 9&0&2\\
\bf(5)&0&0&0&3&4&12&0&3\\
\bf(6)&0&0&0&4&6&16&0&4\\
\end{tabular}}}
\caption[]{The set of solutions of the system (\ref{L/2}), (\ref{L}), (\ref{2b1}), (\ref{2}).}\label{solutions}
\end{table}
The Feynman diagrams corresponding to consecutive cases are
depicted in figure \ref{fig:Diagrams}.

The corresponding vertices assume the following form
\begin{equation}
    V_1=-\frac{1}{2}\sigma_a^{AB}\sigma_b^{CD}\epsilon_{AC}
    \Omega_{IJBD}g^{\mu\nu}\chi_\mu^I\eta^J\partial_\nu
    \varphi^a\varphi^b,
    \nonumber
\end{equation}
\begin{equation}
    V_2=-\frac{1}{4}g^{-1/2}\sigma_a^{AB}\sigma_b^{CD}
    \epsilon_{AC}\Omega_{IJBD}\epsilon^{\mu\nu\lambda}
    \chi_\mu^I\chi_\nu^J\partial_\lambda\varphi^a\varphi^b,
    \nonumber
\end{equation}
and
\begin{equation}
    V_3=\frac{1}{6}g^{-1/2}\epsilon^{\mu\nu\lambda}
    \Omega_{ABCD}\chi_\mu^A\chi_\nu^B\chi_\lambda^C\eta^D.
    \nonumber
    \end{equation}
%
%
The vertex $V_0$ is absent in figure 2, whereas $V_3'$ immediately
vanishes in the diagram {\bf(4)} (see below). It appears that all the Feynman diagrams
but one, i.e. {\bf($\bf3_1$)}, give zero contributions to $Z_I(M)$ according to the following arguments:
%
%
\begin{description}
    \item[(1)]
    $V_3$ vanishes because it contains the exterior product of 3 (harmonic) 1-forms whereas there are only 2 linearly independent ones for $b_1(M)=2$;
    \item[(2)]
    $V_2$ vanishes because it contains the exterior derivative of harmonic forms;
    \item[(3$_1$)]
    This contribution is not zero, and it will be evaluated later on;
    \item[(3$_2$)]
    Analogous to (2);
    \item[(4)]
    It vanishes because one boson external line remains unpaired;
    \item[(5)]
    $V_3$ vanishes because the propagator $\left<\eta\chi\right>$ introduces the exterior derivative for harmonic forms attached to the vertex;
    \item[(6)]
    It vanishes because of unpaired $\eta$ lines (there is no non-zero $\left<\eta\eta\right>$ propagator).
\end{description}

\section{Topological invariant}
The only surviving classical contribution to the partition function (\ref{Z_X(M)})
beyond one loop is given by the integral corresponding to the
diagram ({\bf3}$_1$). This term is analogous to the one appearing in \cite{Rozansky:1997bq}
\begin{eqnarray}
    I(M)&=&\int_{M\times M} \epsilon^{\mu_1\mu_2\mu_3}
    \epsilon^{\nu_1\nu_2\nu_3}\omega_{\mu_1}^{(1)}(x_1)
    \omega_{\nu_1}^{(1)}(x_2)\omega_{\mu_2}^{(2)}(x_1)
    \omega_{\nu_2}^{(2)}(x_2)
    \nonumber\\
    &&\quad G_{\mu_3\nu_3}'(x_1,x_2)\,d^3x_1d^3x_2,
    \nonumber
\end{eqnarray}
where $\omega_\mu^{(1,2)}$ are the basic integral 1-forms.
It gives rise to the Massey product or (Poincare dual) Milnor
linking number
\begin{equation}
    I(M)=\int_M g\wedge dg,
\end{equation}
where $dg=\omega^{(1)}\wedge\omega^{(2)}$. Thus
\begin{equation}
    Z(M)=b(\tilde{X})\tau_R^{-2}(M)I(M),
\end{equation}
with the proportionality coefficient
\begin{eqnarray}
    b(\tilde{X})&
    =&\frac{1}{8\pi^2}
    \int_{\tilde{X}}\sqrt{\tilde{g}}\,d^3\varphi_o'
    \epsilon^{I_1J_1}\epsilon^{I_2J_2}\epsilon^{I_3J_3}
    \epsilon^{I_4J_4}\Omega_{I_1I_2I_3I_4}
    \Omega_{J_1J_2J_3J_4}
    \nonumber
\end{eqnarray}
depending only on the geometry of the manifold $\tilde{X}$.

\section{Finishing remarks}
Our conclusion that the only 3-manifolds $M$ for which the invariant can be non-zero are those with $b_1(M)=2$ could seem to be in contrast with the results of the RW model, where one finds non-zero results for all $b_1(M)\leq3$. But one should take into account that our target space $X$ has an isometry thus is a special case of the more general $X$ considered in \cite{Rozansky:1997bq}.

We would also like to stress that our topological result is not an {\it a posteriori} confirmation of the topological nature of the dualized model but only a ``classical'', i.e.\ the lowest in $\hbar$-expansion, part of the whole partition function $Z(M)$.

\section{Acknowledgements}
The author would like to thank the organizers of the {\it EuroConference on Partial Differential Equations and their Applications to Geometry \& Physics} in Castelvecchio Pascoli
for their kind invitation to actively participate in the conference. The paper is supported by the KBN grant 5 P03B 072 21.

%

\bibliography{moje,TQFT}

\end{document}